\documentclass[submission,copyright,creativecommons]{eptcs}
\providecommand{\event}{HCVS 2021} %
\usepackage{breakurl}             %
\usepackage{amsmath}
\usepackage{amssymb}
\usepackage{graphicx}
\usepackage{tikz}
\usetikzlibrary{automata, positioning, arrows}
\usepackage[final]{commenting} %
\usepackage{paralist} %

\usepackage{listings}
\usepackage{doi}
\usepackage{array}

\newcommand{\id}[1]{{\mathsf{#1}}}
\newcommand{\tuplevar}[1]{\mathbf{#1}}
\newcommand{\tuple}[1]{\ensuremath{\langle #1 \rangle}}
\newcommand{\theory}{\mathbb{T}}
\newcommand{\z}{\ensuremath{\mathbb{Z}}}

\newcommand{\ignore}[1]{}

\newcommand{\costr}[1]{{\mathsf{#1}}}
\newcommand{\project}[2]{\ensuremath{\mathsf{project}}_{#1}(#2)}
\newcommand{\vars}[1]{\mathtt{vars}(#1)} %

\newcommand{\rd}[1]{\ensuremath{\mathtt{rd}(#1)}}

\newcommand{\bk}[1]%
{\textbf{bk~}$\blacktriangleright$\textcolor{red}{#1}$\blacktriangleleft$}
\newcommand{\plg}[1]%
{\textbf{plg~}$\blacktriangleright$\textcolor{blue}{#1}$\blacktriangleleft$}
\newcommand{\jg}[1]%
{\textbf{jg~}$\blacktriangleright$\textcolor{magenta}{#1}$\blacktriangleleft$}
\newcommand{\mh}[1]%
{\textbf{mh~}$\blacktriangleright$\textcolor{orange}{#1}$\blacktriangleleft$}

\newtheorem{definition}{Definition}
\newtheorem{theorem}{Theorem}
\newtheorem{lemma}{Lemma}

\newtheorem{example}{Example}

\usepackage{algorithm}
\usepackage[noend]{algpseudocode}
\algnewcommand\algorithmicforeach{\textbf{for each}}
\algdef{S}[FOR]{ForEach}[1]{\algorithmicforeach\ #1\ \algorithmicdo}

\newcommand{\pcode}[2][\codesize]{
    \fbox{
    \begin{minipage}{0.45\linewidth}
    #1
    \begin{tabbing}
    xx \= xx \= xx \= xx \= xx \= xx \= xx \= xx \= xx \= xx \= \kill
    #2
    \end{tabbing}
    \end{minipage}
    }
  }

\newcommand{\true}{\mathsf{true}}
\newcommand{\false}{\mathsf{false}}

\declareauthor{bk}{Bish}{red}
\declareauthor{jg}{John}{blue}
\declareauthor{pl}{Pedro}{red!50}
\declareauthor{mh}{Manuel}{blue!50}
\declareauthor{jf}{Jose}{green}

\title{Regular Path Clauses and Their Application in Solving Loops} %

\author{Bishoksan Kafle\thanks{Email. \texttt{bishoksan.kafle@imdea.org}}
\institute{IMDEA Software Institute, Spain}   \and
John P. Gallagher\institute{Roskilde University, Denmark}
\institute{IMDEA Software Institute, Spain}
\and
Manuel V. Hermenegildo\institute{IMDEA Software Institute, Spain}
\institute{U. Polit\'{e}cnica de Madrid, Spain}
\and  Maximiliano Klemen\institute{IMDEA Software Institute, Spain}
\and
Pedro L\'opez-Garc\'ia\institute{IMDEA Software Institute, Spain}
\institute{Spanish Council for Sci.\ Research (CSIC)}
\and Jos\'e F. Morales
\institute{IMDEA Software Institute, Spain}
\institute{U. Polit\'{e}cnica de Madrid, Spain}
}

\def\titlerunning{Regular Path Clauses and Their Application in Solving Loops}
\def\authorrunning{Kafle, Gallagher, Hermenegildo, Klemen, L\'opez-Garc\'ia and Morales}
\begin{document}
\maketitle

\begin{abstract}

%
% start input ./abstract.tex

A well-established approach to reasoning about loops during program
analysis is to capture the effect of a loop by extracting recurrences
from the loop; these express relationships between the values of
variables, or program properties such as cost, on successive loop
iterations. Recurrence solvers are capable of computing closed forms for
some recurrences, thus deriving precise relationships capturing the
complete loop execution. However, many recurrences extracted from loops
cannot be solved, due to their having multiple recursive cases or
multiple arguments. In the literature, several techniques for
approximating the solution of unsolvable recurrences have been proposed.
The approach presented in this paper is to define transformations based
on regular path expressions and loop counters that (i) transform
multi-path loops to single-path loops, giving rise to recurrences with a
single recursive case, and (ii) transform multi-argument recurrences to
single-argument recurrences, thus enabling the use of recurrence solvers
on the transformed recurrences.  Using this approach, precise solutions
can sometimes be obtained that are not obtained by approximation
methods.
 
\noindent
\textbf{Keywords}: 
Horn clauses,
path programs,
multi-path recurrences,
multi-argument recurrences.

%
%
%
%
 % end input ./abstract.tex
 
\end{abstract}

%
% start input ./intro.tex
\section{Introduction}
\label{sec:intro}

Inferring the effect of loops is a critical program analysis task, as
loops can give rise to an infinite number of program states, thus
necessitating approximate solutions in general. One approach, often
used in automatic resource analysis, is to extract recurrence
relations from loops, and then try to solve the recurrences to get a
closed form expression \cite{Wegbreit74,granularity,caslog,resource-iclp07,AlbertGM13}. This approach has also been used 
for non-linear invariant synthesis in a series of papers \cite{farzanfmcad15,kincaidpopl18,kincaidpopl19} by Kincaid et al.\ and Humenberger et al.\ \cite{Humenbergervmcai18}. When the recurrences are solvable, precise solutions are obtained. By contrast, approximation methods such as abstract interpretation over some abstract domain
construct solutions whose precision is limited by the expressiveness
of the domain.

We present an approach to solving such loops formulated in terms of
constrained Horn clauses (CHCs), which are capable of expressing the
semantics of imperative programs
~\cite{%
  Peralta-Gallagher-Saglam-SAS98,%
  HGScam06-short,decomp-oo-prolog-lopstr07-short,%
  mod-decomp-jist09-short,GrebenshchikovLPR12,%
  GurfinkelKKN15-short,AngelisFPP15-short,KahsaiRSS16-short,%
  big-small-step-vpt2020-short}. %
Clauses are assumed to have
numerical variables (of infinite precision); we assume that variables values have been abstracted
with respect to their numerical sizes, by performing a program
transformation that uses the desired metrics (such as list length, term
depth, term size, etc.~\cite{resource-iclp07}).

Existing computer algebra systems (CASs) are powerful tools that can
obtain an exact closed-form solution for many mathematical recurrences,
usually deterministic single-argument recurrences with a single recursive
case.  However, many loops give rise to recurrences that are not of this
form and are unsolvable by CASs. That is, they contain multiple
recursive cases, or define functions with multiple arguments, or both.

\paragraph{On multiple recursive cases.} Our approach differs 
from previous works \cite{farzanfmcad15,kincaidpopl18,kincaidpopl19,AlbertGM13} on treating multi-path loops 
that give rise to multiple recursive cases, in that 
we transform each multi-path loop into a program containing nested loops, each of which
has a single path. The transformation preserves all the different paths
and uses regular path expressions. Previous approaches generate one recursive equation for each multi-path
loop by merging different loop paths, which may result in a significant loss of precision.  
Various forms of \emph{control-flow refinement} \cite{GulwaniJK09,Sharmacav11} can transform a 
restricted form of multi-path loops, 
called multi-phase loops, into a semantically equivalent sequence of single path loops.
Polyvariant specialisation~\cite{DomenechGG19,mcctr-fixpt,spec-jlp} can achieve the same goal.
These specialisation techniques are in any case orthogonal and complementary to the ones that we propose. 

As regards solving, Kincaid et al.~\cite{farzanfmcad15,kincaidpopl18,kincaidpopl19} can obtain a closed-form 
solution of a system of inequations, whereas, like \cite{AlbertGM13}, we rely on existing CASs  
to obtain a closed-form solution of a system of recurrences. Humenberger et al.\ \cite{Humenbergervmcai18} can generate polynomial invariants from multi-path numeric loops with polynomial assignments (also known as extended p-solvable loops). They take each single path loop separately, derive a system of recurrences from it and solve them to a closed-form. Then, from the closed-form, they derive the polynomial invariant ideal and combine the ideals of each loop-path iteratively until a fixed-point is reached. In contrast to this,  we do not need a fixed point computation that exploits the special properties of polynomial ideals, but rather aim to transform multi-path loops into a form to which existing solvers can be applied.

\paragraph{On multiple-argument functions.}  Our approach in this area
is motivated by and adapted from the work of Farzan and Kincaid~\cite{farzanfmcad15}. 
In their work, separate recurrences are derived by constructing, for each
argument, a function of one argument, 
namely a loop counter $k$, which defines the  
 value of the argument on the $k^{th}$ loop iteration.
We go beyond \cite{farzanfmcad15}  and present a systematic approach for constructing these functions,
based on regular path programs, and show that the
separate recurrences can be solved to yield a solution for the original multi-argument recurrence.
In~\cite{AlbertGM13} the recurrence is re-expressed as a function of a
single loop counter variable, but separate functions are not derived
for each argument. Farzan and Kincaid~\cite{farzanfmcad15} use a 
quantifier elimination procedure available only to a handful of theories to eliminate the counter,
whereas we follow the approach of Albert et al.~\cite{AlbertGM13} in using a ranking function to estimate its value.

In both of the aspects above, the concept of a path program plays an
essential role in our approach.  The main contributions of the paper
are as follows:

\begin{itemize}
\item A transformation that reformulates multi-path loops (that give
 rise to recurrences with multiple recursive cases) as single-path
 loops (that give rise to recurrences with single recursive case) (\S \ref{sec:path_prog}).

\item A transformation that derives   single argument recurrences from multi-argument ones (\S \ref{sec:path_counters}).
\end{itemize}
The paper is organized as follows. After covering introductory materials in \S \ref{sec:prelim}, we introduce regular path programs in \S \ref{sec:path_prog} and path programs with counters in \S
\ref{sec:path_counters}. Finally, we present concluding remarks  in \S \ref{sec:conclusion}.

 % end input ./intro.tex
 
%
% start input ./prelim.tex
\section{Preliminaries}
\label{sec:prelim}

We assume that a program $P$ is represented as a set of constrained Horn clauses 
(CHCs) of
the form $\id{c}: p(\tuplevar{x}) \leftarrow \phi \wedge
p_1(\tuplevar{x_i}) \wedge \ldots \wedge p_n(\tuplevar{x_n}), n\geq
0$, where $\id{c}$ is a unique identifier for the clause in $P$, $p$ and $p_i$
are predicates, $\tuplevar{x}$ and $\tuplevar{x_i}$ are sequences
of distinct variables (the symbol $\tuplevar{x}$ represents a sequence
of variables $x_1, \ldots, x_n$), and the formula $\phi$ is a conjunction
of constraints over some constraint theory $\theory$. Sometimes we
treat a conjunction of constraints as a set of
constraints. $p(\tuplevar{x})$ is called the \emph{head} and $\phi
\wedge p_1(\tuplevar{x_i}) \wedge \ldots \wedge p_n(\tuplevar{x_n})$
is called the \emph{body} of the clause. 
If $n=0$, then the clause is
called a \emph{constrained fact}, if $n \le 1$ then it is called a
\emph{linear} clause and \emph{non-linear} if it is not linear.  
For convenience we sometimes write an empty conjunction as $\true$,
thus a constrained fact $p(\tuplevar{x}) \leftarrow \phi$ is sometimes written as
as $p(\tuplevar{x}) \leftarrow \phi,\true$; 
and the head of the clause can be
the predicate $\false$.
A program is
called linear if it contains only linear clauses. Normally, we 
write a clause as $p(\tuplevar{x}) \leftarrow \phi,
p_1(\tuplevar{x_i}), \ldots, p_n(\tuplevar{x_n})$ using comma instead
of $\wedge$.
The expression \emph{vars(t)} represents the set of variables of a
term $t$.

A finite \emph{leftmost-derivation} (or simply \emph{derivation}) in a program 
$P$ is a sequence $\leftarrow G_0,\ldots, \leftarrow G_n$ where 
for $0 \le i \le n-1$ $G_i=\varphi_i,q_1(\tuplevar{y_1}), \ldots, q_k(\tuplevar{y_k})$, 
$P$ contains a (suitably renamed) clause
$q_1(\tuplevar{y_1}) \leftarrow \phi,r_1(\tuplevar{z_1}), \ldots, r_m(\tuplevar{z_m})$, 
$\phi_{i+1}=\varphi_i \wedge \phi$ and
 $G_{i+1}=\varphi_{i+1},r_1(\tuplevar{z_1}), \ldots, r_m(\tuplevar{z_m}),q_2(\tuplevar{y_2}), \ldots, q_k(\tuplevar{y_k})$. 
 The derivation is \emph{feasible} if $\varphi_n$ is satisfiable.  
 A derivation is \emph{successful} if it is feasible and $G_n = \varphi_n,\true$.

\begin{figure}[h]
\begin{center}
\vspace{-1em}
\begin{tabular}{m{3cm}m{4cm}m{3cm}}
    \pcode[\small]{
     ﻿$\mathtt{c_1.~ wh(a,b) \leftarrow a>0,b>0,wh(a,b-1)}$ \\
     ﻿$\mathtt{c_2.~ wh(a,b) \leftarrow a>0,b \leq 0, wh(a-1,b+a)}$ \\
     ﻿$\mathtt{c_3.~ wh(a,b) \leftarrow a \leq 0}$
    }
&& 
\begin{tikzpicture}
\node[state] (q1) {$\mathtt{wh}$};
\node[state, right of=q1, xshift=1cm] (q2) {$\mathtt{true}$};
\draw (q1) edge[loop above] node{$\mathtt{c_1}$} (q1);
\draw (q1) edge[loop below] node{$\mathtt{c_2}$} (q1);
\draw[->] (q1) edge node[above]{$\mathtt{c_3}$} (q2);
\end{tikzpicture}
\end{tabular}
\vspace{-2em}
\end{center}
\caption{(left) Multi-path loop and (right) its CFG.  \label{exprogram}}
\end{figure}
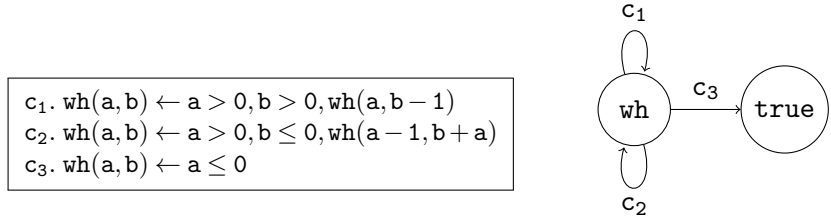

\begin{definition}[Control flow graph, path, loop]
\label{def:cfg}
A control flow graph (CFG) of a linear program $P$ is a labelled
directed graph $\langle N,E\rangle$ where

\begin{itemize}
\item $N$ is a set of nodes consisting of predicates of $P$ (including $\true$ and $\false$);
\item $E$ is a set of labelled edges. There is an edge from $p$ to
  $p_1$ labelled by $\id{c}$ if there is a clause 
  $p(\tuplevar{x}) \leftarrow \phi \wedge p_1(\tuplevar{x_1}) \in P$ having identifier $\id{c}$.
\end{itemize}

A \emph{path} of length $n$ ($ n \ge 0$) in a CFG is a possibly empty sequence of 
$n$ connected edges, and is written as a sequence of \emph{edge labels}.
A \emph{loop} is a non-empty path which starts and ends with the same node $n$ and
does not visit $n$ in between. A loop is \emph{directly recursive} if
it is of length 1. 

\end{definition}

Figure \ref{exprogram} shows an example of a linear program and its CFG.

Let $c_1\ldots c_n$ ($n\ge 0$) be a (possibly empty) path in the CFG for $P$.  
Then the derivation $\leftarrow G_0,\ldots, \leftarrow G_n$ \emph{corresponds} to $c_1 \ldots c_n$ if (i)
$G_0=\varphi_0,p_0(\tuplevar{x_0})$, where $\varphi_0$ is empty and (if $n>0$) the edge $c_1$ starts at $p_0$; 
(ii)  for $1 \le i \le n$, 
$G_{i-1}=\varphi_{i-1},p_i(\tuplevar{x_i})$ and $c_i$ is the identifier of a (renamed) clause 
$p_{i-1}(\tuplevar{x_{i-1}}) \leftarrow \phi_i,p_{i}(\tuplevar{x_i})$ in $P$, 
$\varphi_{i}=\varphi_{i-1} \wedge \phi_i$ and $G_{i}=\varphi_{i},p_{i}(\tuplevar{x_i})$. 
 If $n=0$ then $p_0$ is an arbitrary predicate from $P$.

We say that a path is feasible or successful if and only if the derivation corresponding to the
path is feasible or successful respectively.

\begin{definition}[Recurrence equation] 
\label{def:cr}
A recurrence equation is an equation of the form
 $
 \costr{f}(\tuplevar{x}) = e(\tuplevar{x}) + \sum_{i=1}^n a_i*\costr{f_i}(\tuplevar{x_i}), \phi
$  
where $e(\tuplevar{x})$ is an arbitrary arithmetic expression, $a_i$
some constant, $\costr{f}$ and $\costr{f_i}$ are function symbols, where $\costr{f_i}$ is not necessarily different from $\costr{f}$, $\tuplevar{x}, \tuplevar{x_i}$ are sequence of variables and
$\phi$ is a formula over $\z$. If $n \leq 1$, then it is called a linear (recurrence) equation. 
\end{definition}
A set of such equations is called  a (recurrence) equation system. Each equation in a system is given an id which uniquely identifies it.
A linear equation system is the one in which all equations are linear.

\begin{definition}[Equation graph]
\label{def:eg}
An  equation graph (EG) of a linear equation system $S$ is a labelled
directed graph $G_S= \langle N,E,entry,exit \rangle$ where

\begin{itemize}
\item $N$ is a set of nodes consisting of functions of $S$,
\item $E$ is a set of edges. 
There is an edge from $f$ to
  $f_1$ labelled by $\id{eq}$ if there is an equation 
  $ (\costr{f}(\tuplevar{x}) = e(\tuplevar{x}) +  a_1*\costr{f_1}(\tuplevar{x_1}), \phi) \in S$ having identifier $\id{eq}$.
\item entry and exit  respectively represent the entry and the exit node of $G_S$.
\end{itemize}
\end{definition}

\section{Regular path clauses}
\label{sec:path_prog}

In this section we introduce \textit{path} predicates and clauses;
given a set of linear CHCs $P$, the meaning of the predicate
$\textit{path}_e(p(\tuplevar{x}),q(\tuplevar{x'}))$, where $e$ is a
regular expression describing paths in the CFG for $P$, is that (i) 
there is a feasible path
in the CFG for $P$ starting from the node $p$ and ending at the node $q$, such
that the path is contained in the set described by $e$ and (ii)
$p(\tuplevar{x})$ and $q(\tuplevar{x'})$ are the atoms at the start
and end of the derivation
  corresponding to the path.
(We abuse notation and overload $p$ and $q$ as both function and predicate
symbols).  A well known algorithm by Tarjan~\cite{tarjan81} computes
a regular path expression describing all paths in the CFG from a
designated entry node to an exit node.  We previously developed an
interpreter that is guided by a regular path expression~\cite{big-small-step-vpt2020-short},
and partially evaluated it with respect to a
set of linear CHCs $P$. Here, we outline the idea behind the
interpreter, which is that given a regular expression $e$, each
subexpression of $e$ defines a set of subpaths, which are composed to
give the overall paths. These path predicates are defined by
non-linear CHCs even though the CFG is defined from a set of linear
CHCs.

Regular expressions $e$ over some alphabet $\Sigma$ are expressions of
the form $e ::= c ~\vert~ \epsilon ~\vert~ \emptyset ~\vert~ e_1 e_2
~\vert~ e_1 + e_2 ~\vert~ e^*$, where $c \in \Sigma$. $\Sigma$ in this
case is the set of clause identifiers in $P$. Given a set of linear
CHCs $P$, construct its CFG.  Compute a regular path expression from
the entry node to the exit node of the CFG.  Recall that for any
regular expression $e$, we can compute the set $\textit{first}(e)$,
which is the set of elements of $\Sigma$ that can start a path
described by $e$.  An element of $\textit{first}(e)$ is thus an edge
in the CFG, corresponding to a clause in $P$. The function
$\textit{firstpred}(e)$ returns the set of predicates in the head of
some element of $\textit{first}(e)$. Figure \ref{fig:regpathprog}
shows how path clauses are constructed based on the structure of the
regular expressions.  Note that we choose to define the clauses for an
expression $e^*$ using left recursion, that is, corresponding to the
expansion $e^* = \epsilon + e^*e$, rather than the equivalent $e^* =
\epsilon + ee^*$, which might be expected. This enables the generation
of suitable recurrence equations, which will be discussed in Section \ref{sec:path_counters}.
 
 \begin{figure}
 \vspace{-1em}
 \[
\begin{array}{|l|l|}
\hline
c &  \textit{path}_c(p(\tuplevar{x}),q(\tuplevar{x'})) \leftarrow \phi., \text{where clause }  p(\tuplevar{x}) \leftarrow \phi, q(\tuplevar{x'})
\in P \text{ has identifier } c\\
\hline
\epsilon & \textit{path}_{\epsilon}(p(\tuplevar{x}),p(\tuplevar{x})) \leftarrow \true.\\
\hline
\emptyset & \text{no clause}\\
\hline
e_1 e_2 & \textit{path}_{e_1 e_2}(p(\tuplevar{x}),z) \leftarrow \textit{path}_{e_1}(p(\tuplevar{x}),q(\tuplevar{x'})),\textit{path}_{e_2}(q(\tuplevar{x'}),z)., \text{ for each } q \in \textit{firstpred}(e_2)\\
\hline
e_1 + e_2 &  \textit{path}_{e_1 + e_2}(p(\tuplevar{x}),z) \leftarrow \textit{path}_{e_1}(p(\tuplevar{x}),z).\\ 
& \textit{path}_{e_1 + e_2}(p(\tuplevar{x}),z) \leftarrow \textit{path}_{e_2}(p(\tuplevar{x}),z).\\
\hline
e^* &  \textit{path}_{e^*}(p(\tuplevar{x}),p(\tuplevar{x})) \leftarrow \true.\\
& \textit{path}_{e^*}(p(\tuplevar{x}),p(\tuplevar{x}'')) \leftarrow \textit{path}_{e^*}(p(\tuplevar{x}),p(\tuplevar{x'})),\textit{path}_{e}(p(\tuplevar{x'}),p(\tuplevar{x''})).\\
\hline
\end{array}
\]
\vspace{-2em}
\caption{Formation of path clauses corresponding to a regular
  expression $e$, starting from node $p$.}\label{fig:regpathprog}
\end{figure}

\begin{theorem}\label{prop2}
Let $P$ be a set of linear CHCs, $G$ be its CFG and let $p$ and $q$ be
(possibly identical) predicates in $P$ (including $\false$ and
$\true$).
Let $e$ be a regular expression describing all paths from $p$ to $q$
in the CFG. Then there is a feasible 
derivation 
$\leftarrow p(\tuplevar{x}), \ldots, \leftarrow
\varphi_n,q(\tuplevar{x'})$ in $P$ if and only if
there is a successful derivation 
  $\leftarrow \textit{path}_e(p(\tuplevar{x}),q(\tuplevar{x'})),
  \ldots, \leftarrow\varphi_n,\true$ in the path clauses for $P$ and $e$.
  \end{theorem}
In other words, the path clauses capture all feasible derivations of the original set of clauses $P$. 
It should be emphasised that the path clauses capture derivations in $P$, as stated by Theorem \ref{prop2},
but the path clauses themselves are not intended as an executable program. Clearly the left-recursive form
is not amenable to execution with the standard computational strategy, and the epsilon steps after loops 
are likely to introduce non-determinism.  So is not the intention to analyse the computational cost of
the path program or its termination behaviour.
The only property of the path clauses that we exploit in Section \ref{sec:path_counters} is that $\textit{path}$
predicates capture the relationship between the values of variables at the start and end of derivations in $P$. The notion of path clauses is related to the big-step semantics \cite{big-small-step-vpt2020-short} and path predicaes are sometimes called summary predicates. 

\subsection{Eliminating multi-path loops}\label{sec:multipath}

Theorem \ref{prop2} holds for any expression $e$ that describes all paths from $p$ to $q$
in the CFG, thus we can freely transform $e$ into an equivalent expression $e'$ and 
perform the generation of the path clauses with respect to $e'$.
A regular expression  of the form $(e_1+ \cdots + e_m)^*$ is called a \emph{multi-path-loop expression}, 
where $e_1,\ldots, e_m$ are the
paths through the loop.  Any regular expression can be transformed to an equivalent regular expression 
not containing any
multi-path-loop expression, by repeatedly replacing any subexpression of the form $(e_1+ e_2)^*$ by 
an equivalent regular expression
$(e_1^* e_2^*)^*$,
$e_1^*(e_2 e_1^*)^*$,  or $e_2^*(e_1 e_2^*)^*$, among others.  We discuss the choice of replacement expression 
in Section \ref{sec:conclusion}.

An algorithm for removing multi-path loops using regular expression transformation is given in Alg.\ \ref{alg:mploops}. Given a set of linear clauses with its CFG and distinguished entry and exit nodes, it first computes regular expression describing all paths from entry to the exit node using using Tarjan's algorithm~\cite{tarjan81}. The expression is then rewritten without the choice (+) operator which is then used to transform the original program based on Figure \ref{fig:regpathprog}. The resulting program does not contain multi-path loops.
\begin{algorithm}
   \caption{Algorithm for removing multi-path loops \label{alg:mploops}}
    \begin{algorithmic}[1] 
	\State \textbf{Input}: CFG $G$ of linear clauses $P$, entry node $p$ and exit node $q$
	\State \textbf{Output}: Program $P'$ without multi-path loops
	\State $e \gets reg\_path\_expr(G,p,q)$ \Comment{Compute regular path expression \cite{tarjan81} }
\State $e \gets rewrite\_wo\_choice(e)$ \Comment{Rewrite $e$ without choice operator within star}
\State $P' \gets construct\_path\_cls(P, e, p)$ \Comment{Figure \ref{fig:regpathprog}}
\State \Return $P'$
    \end{algorithmic}
\end{algorithm}
\vspace{-1em}

\begin{theorem}
The output of Algorithm \ref{alg:mploops} is a program that contains no multi-path loop. 
\end{theorem}

\begin{figure}[h]
\vspace{-1em}
\begin{center}
\begin{tabular}{c}
   
    \pcode[\small]{
$\mathtt{path(wh(a,b),\true) \leftarrow
   wh_2(a,b,a',b'),
  wh_5(a',b',a'',b''),
   ~a'' \le 0.}$\\
$\mathtt{wh_2(a,b,a,b) \leftarrow
  \true.}$\\
$\mathtt{wh_2(a,b,a',b'-1) \leftarrow
   wh_2(a,b,a',b'),
   ~a'>0,
   ~b'>0.}$\\
$\mathtt{wh_5(a,b,a,b) \leftarrow
  \true.}$\\
$\mathtt{wh_5(a,b,a'',b'') \leftarrow
   wh_5(a,b,a',b'),
  ~a'>0,
   ~b' \le 0,
   wh_2(a'-1,b'+a',a'',b'').}$
    }
\end{tabular}
\end{center}
\vspace{-1em}
\caption{The path clauses for Fig.\ \ref{exprogram} (left) wrt  $c_1^*(c_2 c_1^*)^* c_3$ (equivalent  to $(c_1+ c_2)^*c_3$).  \label{fig:pathprog}}
\end{figure}

\begin{example}\label{ex:pathprog}

Let $P$ be the set of clauses in Figure \ref{exprogram} (left).  A regular path expression for paths in its 
CFG from $wh$ to $\true$ is $(c_1 + c_2)^* c_3$.  An equivalent path expression with no multi-path-loop expressions 
is $c_1^*(c_2 c_1^*)^* c_3$.
Figure \ref{fig:pathprog} shows the path clauses for $P$ derived using the latter expression, after some path predicates 
have been unfolded and the following renamings applied: $wh_2(a,b,a',b')$ is a renaming of 
$\textit{path}_{c_1^*}(wh(a,b),wh(a',b'))$ and $wh_5(a,b,a',b')$ is a renaming of
$\textit{path}_{(c_2 c_1^*)^*}(wh(a,b),wh(a',b'))$.  

\end{example}

Thus we can replace the original problem of analysing a multi-path loop by the problem of
analysing a program with nested single-path loops. In the example, the predicate $wh_2$ is nested within $wh_5$.  
Furthermore, each loop predicate has multiple input and output values.  In the next section,
we investigate how to analyse the relationship between the input and 
output values of each loop.

 % end input ./path_program.tex
 
%
% start input ./multiargs.tex
\section{Regular path clauses with counters}
\label{sec:path_counters}

In Section \ref{sec:path_prog} it was shown that multi-path loops
could be eliminated in path clauses, by transforming the regular path
expression for the loop.
Thus, all loops after the transformation are represented by a single directly recursive 
path clause of the form:
$\mathtt{path_{e^*}(\tuplevar{x},\tuplevar{x_2})} \leftarrow \mathtt{path_{e^*}(\tuplevar{x},\tuplevar{x_1})},\mathtt{path_{e}(\tuplevar{x_1},\tuplevar{x_2})}.$
Here $\mathtt{path_{e}(\tuplevar{x_1},\tuplevar{x_2})}$ is the path for the loop body, which may itself contain loops, but
these do not depend on the recursive predicate $\mathtt{path_{e^*}}$ and can be solved separately; let us say that the
solution is an expression $\phi(\tuplevar{x_1},\tuplevar{x_2})$.  Then the loop to be solved is a single directly recursive clause 
$\mathtt{path_{e^*}(\tuplevar{x},\tuplevar{x_2})} \leftarrow \phi(\tuplevar{x_1},\tuplevar{x_2}),\mathtt{path_{e^*}(\tuplevar{x},\tuplevar{x_1})}.$
Due to the left-recursive form of the  path, the input arguments $\tuplevar{x}$ remain constant from one recursive call
to the next;  this is an important property when forming recurrence equations, as will be seen.

We now introduce a counter $\mathtt{k}$ to such clauses,
representing the length of the loop path. This allows us to capture the effect of the loop 
after $\mathtt{k}$ iterations.
After adding 
the counter, we obtain the following clauses for each single-path loop.  $\mathtt{k}$ is considered to be an input  and  is decremented until  it reaches 0.
\[
\vspace{-1em}
\begin{array}{l}
\mathtt{path_{e^*}(k,\tuplevar{x},\tuplevar{x_2})} \leftarrow \mathtt{k>0}, \phi(\tuplevar{x_1},\tuplevar{x_2}),\mathtt{path_{e^*}(k-1,\tuplevar{x},\tuplevar{x_1})}.\\
\mathtt{path_{e^*}(k,\tuplevar{x},\tuplevar{x})} \leftarrow \mathtt{k=0}.
\end{array}
\]
We also assume that there is a \emph{ranking function}~\cite{113433,PodelskiR04b,BagnaraMPZ10TR}
for the loop, guaranteeing termination; that is, 
there is an upper bound on the number 
of iterations $k$.  A ranking function for the loop is a function on the input values such that $r(\tuplevar{x}) \in \mathbb{N}$ and 
$ \phi(\tuplevar{x_1},\tuplevar{x_2}) \rightarrow r(\tuplevar{x_1}) > r(\tuplevar{x_2})$.  
Thus the maximum value of $\mathtt{k}$ in successful derivations starting from
$\mathtt{path_{e^*}(k,\tuplevar{x},\tuplevar{x_2})}$ is $r(\tuplevar{x})$.

\begin{example}\label{ex:conversion_unary_args}
Consider a loop clause $\mathtt{wh(x_1,y_1) \leftarrow
  x_1>0,y_1>0, x_2=x_1-1, y_2=y_1+x_1, wh(x_2,y_2)}$. Starting from the input, $(x_1,y_1)$, we aim  to find 
the values of these variables when the loop terminates, in terms of the original ones when entering the loop.
For this purpose, let us first
define a path of length $k$ going from
$\mathtt{(x,y)}$ to $\mathtt{(x_2,y_2)}$ as a relation
$\mathtt{path(k,x,y,x_2,y_2)}$. The clauses for $\mathtt{path}$ are
as follows.
\begin{equation*}
\vspace{-1em}
\begin{array}{lcl}
\mathtt{path(k,x,y,x_2,y_2)}  &
\mathtt{\leftarrow} &
\mathtt{k>0, k_1=k-1, x_2=x_1-1, y_2=y_1+x_1,} \\
& & \mathtt{path(k_1,x,y,x_1,y_1),} 
 \mathtt{x_1>0,y_1>0.} \\
\mathtt{path(k,x,y,x,y)}  & \mathtt{\leftarrow} & \mathtt{k=0.}
\end{array}
\end{equation*}

The recursive clause can be read as saying: if there is a path of length $k-1$ from
$(x,y)$ to $(x_1,y_1)$, and $x_1>0,y_1>0$, then the path can be
extended on the right to a path of length $k$ from $(x,y)$ to
$(x_2,y_2)$, where $(x_1,y_1,x_2,y_2)$ satisfy the constraints from
the loop clause above.
In other words,  $\mathtt{path(k,x,y,x_2,y_2)}$ means that given $x,y,k$ as inputs, $x_2$ and $y_2$  
are the values of $x$ and $y$ after
$k$ iterations of the $\mathtt{wh}$ loop.
\end{example}

\subsection{Deriving recurrences from path clauses with counters}

Given such a loop clause with a counter, we formulate recurrence equations for each of the variables at
the end of the path (called the \emph{output} variables of the loop).  
This can be done systematically using the method described in
\cite{DBLP:journals/toplas/DebrayL93,resource-iclp07}; here we give an informal account of
the construction.  
Given $\mathtt{path_{e^*}(k,\tuplevar{x},\tuplevar{y})}$ representing a loop predicate $\mathtt{p}$,
where $\tuplevar{x}, \tuplevar{y}$ are $m$-tuples of variables respectively representing the values of variables
at the start and end respectively of the path,
 each of the $m$ variables in $\tuplevar{y}$ is a function of $\mathtt{k,\tuplevar{x}}$. This is due to the fact that
 given $\mathtt{k}$ and $\tuplevar{x}$, the values of the output $\tuplevar{y}$ are 
completely determined.
Hence, we can define $m$ functions $\mathtt{p^1(k,\tuplevar{x})},\ldots,\mathtt{p^m(k,\tuplevar{x})}$,
giving the value of the respective elements of the $m$-tuple $\tuplevar{y}$.
That is $\mathtt{path_{e^*}(k,\tuplevar{x},\tuplevar{y})}$ is equivalent to
$\mathtt{p^1(k,\tuplevar{x})=y_1} \wedge \ldots \wedge \mathtt{p^m(k,\tuplevar{x})=y_m}$.
The recurrence equations are then obtained by substituting this functional expression for the $\mathtt{path_{e^*}}$ 
atoms
in the loop clauses.  

We say that function $f$ \emph{depends on} function $g$ in a system of recurrences $S$ 
if $g$ is needed in order to evaluate $f$.  There is a \emph{cyclic} dependency between $f$ and $g$ if
they depend on each other.  If $S$ contains no cyclic dependencies, then we can send the recurrences 
to a CAS to be solved in topological order.  

\begin{example}
[Continued from Example \ref{ex:conversion_unary_args}]
\label{ex:conversion_unary_args_2}

Define two functions $\mathtt{wh^x}(k,x,y)$ and $\mathtt{wh^y}(k,x,y)$ defining the values of $\mathtt{x}$ and 
$\mathtt{y}$ after $k$ iterations of the $\mathtt{wh}$ loop.  The functions are defined as follows.
\begin{equation}
\label{eq_size_x}
\mathtt{wh^x}(k,x,y)
=
\begin{cases}
 \mathtt{wh^x}(k-1,x,y)-1, ~~~~ for~~~~ k>0, \\
x,  ~~~~ for~~~~ k=0
\end{cases}
\vspace{-1em}
\end{equation}
\begin{equation}
\label{eq_size_y}
\mathtt{wh^y}(k,x,y)=
\begin{cases}
 \mathtt{wh^y}(k-1,x,y)+\mathtt{wh^x}(k-1,x,y), ~~~~ for~~~~ k>0, \\
y,  ~~~~ for~~~~k=0
\end{cases}
\end{equation}
Since there are no cyclic dependencies between $\mathtt{wh^x}$
and $\mathtt{wh^y}$, they can be solved in reverse
topological order of the strongly connected components (SCCs); that is
equations for $\mathtt{wh^x}$ are solved first and then equations for
$\mathtt{wh^y}$, reusing solutions from the lower to the higher
SCC when necessary.
\end{example}

\subsection{Removal of symbolic constant arguments}
Since Equation~\ref{eq_size_x} has multiple arguments, it is not a
\emph{mathematical recurrence} that can directly be solved by the
CASs. Note, however, that the arguments $x,y$ remain constant and
unconstrained throughout all recursions of the equation.
In other words,
$x~\text{and}~y$ play no role in the recursive case of
Equation~\ref{eq_size_x}; $x$ appears in the base-case and will be
present in the solution of the equation.  But they can be removed as
arguments of $\mathtt{wh^x}$ and the occurrence of $x$ in the
base case can be replaced by a constant function returning $x$. The resulting
recurrence equation will have the same solution as the original. 
We call this type of arguments \emph{symbolic constants}, and propose
a method for inferring which arguments can be classified as such in
Section~\ref{sec_sc}.

\begin{example}[Continued from Example \ref{ex:conversion_unary_args_2}]
\label{ex:conversion_unary_args_3}
Making these transformations, we
obtain the simplified Equation~\ref{eq_size_x_simpl}, where $c_x$ is a
constant function representing the argument $x$. We abuse notation and overload $\mathtt{wh^v}$ with different arities throughout the paper.
\begin{equation}
\label{eq_size_x_simpl}
\mathtt{wh^x}(k)
=
\begin{cases}
 \mathtt{wh^x}(k-1)-1, ~~~~ for~~~~ k>0, \\
c_x,  ~~~~ for~~~~ k=0
\end{cases}
\vspace{-0.7em}
\end{equation}
Observe that this equation does satisfy the syntactic form required by the
CASs and can be solved to obtain $\mathtt{wh^x}(k)=c_x-k$ as a
solution. This is also the solution of Equation \ref{eq_size_x}, as we
shall see later (Lemma \ref{lemma_soundness_sc}). Thus, we have $\mathtt{wh^x}(k,x,y)=x-k$. 
Replacing the solution of
$\mathtt{wh^x}(k,x,y)$ in Equation \ref{eq_size_y} and simplifying, we
obtain Equation \ref{eq_size_y_with_x_replaced}:
\begin{equation}
\label{eq_size_y_with_x_replaced}
\mathtt{wh^y}(k,x,y)=
\begin{cases}
 \mathtt{wh^y}(k-1,x,y)+ x-k+1, ~~~~ for~~~~ k>0, \\
y,  ~~~~ for~~~~k=0
\end{cases}
\vspace{-0.7em}
\end{equation}
Now, following similar reasoning, $x~\text{and}~y$ are classified as symbolic
constants. This allows further simplification, generating Equation
\ref{eq_sz_y_t}:

\begin{equation}
\label{eq_sz_y_t}
\mathtt{wh^y}(k)=
\begin{cases}
 \mathtt{wh^y}(k-1)+ c_x-(k-1), ~~~~ for~~~~ k>0, \\
c_y,  ~~~~ for~~~~ k=0
\end{cases}
\end{equation}
which  can be solved to yield $\mathtt{wh^y}(k)= c_y - \frac{1}{2} k (k - 2x -
1)$ and hence $\mathtt{wh^y}(k,x,y)=
y-\frac{1}{2}k^2+kx+\frac{1}{2}k$.
\end{example}

This approach has  advantages and disadvantages: on the one hand (i) identifying
symbolic constants and simplifying equations can be done locally, for
instance, at SCC level (where global analysis is not required) and
(ii)
 when solving equations corresponding to a higher SCC, all
variables except the counter $k$ in lower SCCs become \emph{symbolic
  constants} resulting in equations with an unary argument $k$.

On the other hand, the solution shifts the problem 
to estimating the value of the path counter variable ($k$)
in terms of the original
arguments. Observe that $k$ decreases precisely by 1 in each recursive
call to the function and always stays non-negative.  Since $k$ cannot
always be computed precisely, it needs to be approximated. As mentioned
earlier,
we assume that the loop has a ranking function that  decreases at least by
1 in each recursive call. Let $r$ be the value of the ranking function
in the initial state. Then, $k \in [0,r]$ is a safe bound for $k$.

\begin{example}[Continued from Example \ref{ex:conversion_unary_args_3}]
\label{ex:conversion_unary_args_4}
Going back to the original loop clause in Example  \ref{ex:conversion_unary_args}, 
a ranking function for $\mathtt{wh}(x,y)$ is $x$ (or more precisely $max(0,x)$, but we 
assume that 
$x$ may not have negative values).
Let us say that 
$\mathtt{wh}(x_1,y_1)$  is the call at the start of the loop. Then the loop
executes $k$ times where $k$ is in the interval 
$[0,x_1])$. Substituting this interval for $k$ in the solutions of
Equations~\ref{eq_size_x} and \ref{eq_size_y} and using interval arithmetic,
we can infer that, when
the loop terminates, the values of $x_1$ and $y_1$ lie in intervals
$[0,x_1]$ and $[y_1-1/2x_1^2,y_1+1/2x_1+x_1^2]$ respectively.
\end{example}
The use of interval arithmetic gives sound results but can lose precision, 
in particular since the value substituted for $k$ should be the same for both output
variables $x$ and $y$.  This is a well known source of imprecision in interval arithmetic; for instance, the interval approximation of $x*x$ where $x$ is in the interval $[-1,1]$
is $[-1,1]$ since it includes the result $-1*1$, whereas taking into account that the same value of $x$ is used for each occurrence 
would give the positive interval $[0,1]$ since the sign of both multiplicands is the same. 
However, for the inference
of upper bounds, interval arithmetic is expected to give reasonable results.
Further experimentation will be needed.

\paragraph{Summary of the procedure.}
Based on the discussion above, we outline below the steps for
inferring relations between variables before and after the loop (also
known as loop summaries).  We assume that the loop is defined by a
single recursive linear path clause; that is, multi-path loops have
been eliminated by the technique presented in Section
\ref{sec:path_prog}, and the loop body (defined by predicates in lower
SCCs of the path clauses) has been solved.
\begin{enumerate}

\item Given a loop clause expressed as a single recursive path clause,
add a counter $k$ to the clause, with a base case where $k=0$.

\item Then, given an input tuple $\tuplevar{x}$ and $k$, set up equations
for output tuple $\tuplevar{z}$ using the path clause, using methods such as as in
\cite{DBLP:journals/toplas/DebrayL93}.

\item Transform multi-argument equations to single-argument equations by
  detecting and removing symbolic constants, as discussed in Section \ref{sec_sc}.

\item Solve the resulting equations using CASs and replace the path counter in the
  solution with the value of the ranking function in the initial state
  of the original loop clause.

\end{enumerate}

\subsection{Inference of symbolic constants}
\label{sec_sc}

In the procedure above, multi-argument functions were transformed to
single-argument functions by removing arguments that were symbolic constants.
Symbolic constants can be detected using an adapted data flow
analysis technique based on reaching definitions \cite{popa-nielson},
RD for short. RD provides information about where variables have most
recently obtained their values. We adapt it for a system of linear recurrence equations. 

The algorithm assigns to each node of an EG a set of
pairs of the form $(v,e)$ to express the fact that the variable $v$ may
have been last defined (or constrained) in the equation $e$. We
refer the readers to \cite{popa-nielson} for details on the
analysis. Next, we explain how the central concept of a ``definition" of a variable
is adapted for linear equation systems.

\begin{definition}[Defined and constrained variable]
Given an EG $\langle V,E,entry,exit \rangle$ of a linear equation system, a variable
$v$ is \emph{defined} in the edge $e$, where $e$ is the equation $(\costr{p}(\tuplevar{x}) = expr(\tuplevar{x}) +  a_1*\costr{q}(\tuplevar{x'}), \phi)$
if $\phi \models_\theory v \neq v'$, where $v' \in
\tuplevar{x'}$
and $v \in \tuplevar{x}$ are the corresponding occurrences of an argument in the right and left sides of $e$.
 We denote it by
$(v,e)^d$. Similarly, $v$ is \emph{constrained} in that edge if $v \in
\vars{\project{\tuplevar{x}}{\phi}}$. We denote it by $(v,e)^c$.
\end{definition}

Based on this definition we now define symbolic constants.
\begin{definition}[Symbolic constant]
Given an EG of a linear system of equations $S$, a variable of $S$ is called a symbolic
constant if it is neither defined nor constrained in any edges of the
EG.
\end{definition}

That is, when adapting RD analysis to infer symbolic constants for a
linear system of equations, we consider a
weaker notion of ``defined variable" and say that a variable is defined
(denoted as $(v,e)$) if $(v,e)^d \vee (v,e)^c$.  With this
definition, the standard analysis can be used as follows. Let $G_S$ be
an EG of some equation system $S$ and let $\rd{n}$ denote the results of
the analysis for node $n$.
\begin{itemize}
\item Compute RD assignment for $G_S$, starting with
  $\rd{n}=\emptyset$ for all nodes $n$ of $G_S$. Let $\rd{exit}$ be
  the result for the $exit$ node.
\item Let $V$ be a set of variables of interest. $v \in V$ is a
  symbolic constant if $(v,\_) \not\in \rd{exit}$. In other words,
  no definition of $v$ reaches the exit node. Since it was not defined
  or constrained on entry, we can conclude that it was never defined
  or constrained in $S$.
\end{itemize}

\begin{example}[RD analysis and symbolic constants] 
\label{ex:sc}
Let $G_S=\langle N,E,entry,halt \rangle$ below be the EG of Equation \ref{eq_size_x} with its cases
labelled as edges $e_1$ and $e_2$, where
$N=\{entry, \mathtt{wh^x},  halt\}$ and $E=\{e_0, e_1,e_2\}$, where $e_0$ is the entry edge from $entry$ to
$ \mathtt{wh^x}$.
The result of the analysis of $G_S$ is as
follows. 
\[ \rd{entry} = \emptyset, ~~\rd{\mathtt{wh^x}}= \{(k,e_1)\}, ~~\rd{halt}= \{(k,e_2)\} \] 
It is easy to see that (i) $(k,e_1)^d \wedge (k,e_1)^c$: $k$ is both
defined and constrained in the recursive equation $e_1$;
(ii) $(k,e_2)^c$: $k$ is only constrained in the base case $e_2$; and
(iii) variables are neither defined nor constrained in $e_0$. 
Since $(x,\_ ) \not \in \rd{exit}$ and $(y,\_) \not
\in \rd{exit}$, both $x$ and $y$ are symbolic constants.
\end{example}

\begin{lemma}[Soundness of removing symbolic constants from equations]
\label{lemma_soundness_sc}
Let $S$ be a system of equations and $V$ be a set of symbolic
constants such that $U \subseteq V$ appear as arguments of a function
$f$ in $S$. Let $S'$ be an another system obtained by removing the
variables in $U$ from arguments of $f$ and replacing any other occurrences
of variables in $U$ by constant functions that return the initial value they are assigned
when evaluating a call to $f$. Then $S$ and $S'$ have the
same solution.
\end{lemma}

\subsection{Solving multi-argument recursive  equations}
In previous sections, we showed how to solve loops represented as
clauses using recurrences.  We can apply the same technique to solve some
equations with multi-argument functions
by first transforming them to tail-recursive clauses and solving them. It is not always possible to transform 
all the recursive calls to tail recursive calls but we can rewrite some programs, for example, 
using accumulator(s),  continuation-passing style (introducing a stack) or the 
standard techniques based on unfold-fold transformations \cite{Burstall:77:78} to achieve the required form. 
In this paper, we make use of the latter as it fits well with our approach.

Consider the simple case of recurrence equations of Definition \ref{def:cr}, namely
$ \costr{f}(\tuplevar{x}) = e(\tuplevar{x}) +  \costr{f}(\tuplevar{x_i}), \phi$ or $ \costr{f}(\tuplevar{x}) = e(\tuplevar{x}), \phi$.
Functions of this form can be transformed by unfold-fold to  a tail-recursive \emph{accumulating} form,
which can then be written as CHCs of the required form.  We illustrate this with an example below.
 \begin{equation}
\label{eq:multi-arg}
\mathtt{wh(x_1,y_1)=} 
\begin{cases}\mathtt{x_1+y_1+1 + wh(x_1-1,y_1+1) ~\text{for}~ x_1>0} \\
\mathtt{0 ~\text{for}~ x_1 \leq 0}
\end{cases}
\end{equation}
Following a fold-unfold transformation, we can obtain the following equivalent system of equations.
\begin{equation}
\label{eq:multi-arg-1}
\mathtt{wh(x_1,y_1)=} 
\begin{cases}\mathtt{wh\_aux(x_1-1,y_1+1,x_1+y_1+1) ~\text{for}~ x_1>0} \\
\mathtt{0 ~\text{for}~ x_1 \leq 0}
\end{cases}
\vspace{-1em}
\end{equation}
 \begin{equation}
\label{eq:multi-arg-2}
\mathtt{wh\_aux(x_1,y_1,z_1)=} 
\begin{cases}\mathtt{wh\_aux(x_1-1,y_1+1,x1+y1+1+z_1) ~\text{for}~ x_1>0} \\
\mathtt{z_1 ~\text{for}~ x_1 \leq 0}
\end{cases}
\end{equation}
The recursive function $\mathtt{wh\_aux}$ can now be solved by transforming its first equation into the CHC.
 \[ 
 \begin{array}{ll}
 \mathtt{wh\_aux(x_1,y_1,z_1,w_1) \leftarrow }&\mathtt{x_1>0, x_2=x_1-1, y_2=y_1+1, z_2=x_1+y_1+1 +z_1, w_2=w_1}\\
&\mathtt{wh\_aux(x_2,y_2,z_2,w_2)}. 
 \end{array}
 \vspace{-1em}
 \] 

\noindent
This loop clause has ranking function $x_1$. The corresponding path clauses with counter $k$ is given as:
\[
\begin{array}{ll}
\mathtt{path(x,y,z,w,k,x_2,y_2,z_2,w_2)}  \leftarrow &\mathtt{x_1>0, k>0, k_1=k-1,}\\
&\mathtt{x_2=x_1-1, y_2=y_1+1,} \mathtt{z_2=x_1+y_1+1 +z_1,  w_2=w_1} \\ 
&\mathtt{path(x,y,z,w,k_1,x_1,y_1,z_1,w_2)}.  \\
\mathtt{path(x,y,z,w,k,x,y,z,w)}  \leftarrow &\mathtt{k=0}.
\end{array}
\vspace{-1em}
\]
 Given $x,y,z,w,k$ as inputs, the value of $z_2$ represented as $\mathtt{wh\_aux^z}(x,y,z,w,k)$ is given by the following equation, where we abbreviate $\mathtt{wh\_aux}$ by $\mathtt{f}$.
\begin{equation}
\label{eq_sz_Cost}
\mathtt{f^z}(x,y,z,w,k)=
\begin{cases}
\mathtt{f^x}(x,y,z,w,k-1)+ \mathtt{f^y}(x,y,z, w,k-1)+ 1 + \mathtt{f^z}(x,y,z,w,k-1), ~~~~ \text{for } k>0, \\
z,  ~~~~  \text{for } k=0 
\end{cases}
\vspace{-1em}
\end{equation}
Assume that we have already computed the following:
$\mathtt{f^x}(x,y,z,w,k)=x-k$ and $\mathtt{f^z}(x,y,z,w,k)=y+k$. Reusing
them in Equation \ref{eq_sz_Cost} and simplifying, we obtain Equation
\ref{eq_sz_Cost_simpl}.
\begin{equation}
\label{eq_sz_Cost_simpl}
\mathtt{f^z}(x,y,z,w,k)=
\begin{cases}
 x+y+ 1 + \mathtt{f^z}(x,y,z,w,k-1), ~~~~ for~~~~ k>0, \\
z,  ~~~~ for~~~~ k=0
\end{cases}
\vspace{-1em}
\end{equation}
Noting that $x$, $y$, $z$ and $w$ are symbolic constants,
this can be solved to yield $\mathtt{wh^z}(x,y,z,k)=z+k* (x+y+1)$. Now,
mapping the results back to the original clause, we will have $k \in
[\ 0, x_1]\ $ and the value of $z_1$ after the termination of the loop
in  $[\ z_1, z_1+x_1*(x_1+y_1+1)\ ]$, which is the
solution of Equation \ref{eq:multi-arg-2}.

 % end input ./multiargs.tex
 
%
% start input ./workout_ex.tex
\section{Solving the running example using  interval arithmetic}
\label{sec:ex}

We have now described all the necessary components for solving the running example in Fig.~\ref{exprogram}. We proceed to solve it by solving the corresponding path clauses in Fig.~\ref{fig:pathprog} in the reverse topological order of its SCC graph. That is, we first solve clauses corresponding to  $\mathtt{wh_2}$, and then clauses for $\mathtt{wh_5}$ and finally   $\mathtt{path}$. Solving the clauses 2 and 3 following the procedure described above, we obtain $\mathtt{wh_2(a,b,a',b') \leftarrow a'=a, b' =b-k_1, 0 \leq k_1 \leq b}$ where $\mathtt{k_1}$ is the counter of the loop which is bounded from above by its ranking ranking function $\mathtt{b}$.  Now, clause 5, after reusing the solution of $\mathtt{wh_2(a,b,a',b')}$, becomes a linear clause
\[ \mathtt{wh_5(a,b,a'',b'') \leftarrow
   wh_5(a,b,a',b'),
  ~a'>0,
   ~b' \le 0, a''=a'-1, b''=b'+a'-k_1, 0 \leq k_1 \leq b'+a' %
   .} \] 
Note that the counter variable $k_1$ is not eliminated but rather carried along. Further, note that, its upper bound can be simplfied to $\mathtt{0 \leq k_1 \leq a'}$ since $\mathtt{b' \leq 0}$ yielding a simplified clause:
\[ \mathtt{wh_5(a,b,a'',b'') \leftarrow
   wh_5(a,b,a',b'),
  ~a'>0,
   ~b' \le 0, a''=a'-1, b''=b'+a'-k_1, 0 \leq k_1 \leq a' 
   .} \] 
 Now solving this clause along with clause 4, we get \[ \mathtt{wh_5(a,b,a',b') \leftarrow a'=a-k_2, b' =b + \frac{1}{2}* k_2 (2*a - k_2 - 2*k_1 + 1), 0 \leq k_1 \leq a'+1, 0 \leq k_2 \leq a}.\]
 Finally, we reuse the solution of $\mathtt{wh_2(a,b,a',b')}$ and $\mathtt{wh_5(a,b,a',b')}$ in the $\mathtt{path}$ clause obtaining
 
$ \mathtt{path(wh(a,b),\true) \leftarrow
   a'=a, b' =b-k_3, 0 \leq k_3 \leq b,a''=a'-k_2,}$ \\ %
  $ \mathtt{~~~~~~~~~~~~~~~~~~~~~~~~~~~~~~~~~b'' =b' + \frac{1}{2}* k_2 (2*a' - k_2 - 2*k_1 + 1), 
  0 \leq k_1 \leq a''+1, 0 \leq k_2 \leq a', %
   ~a'' \le 0.}$ \\
 Note that $\mathtt{a''}$ and $\mathtt{b''}$ represent the values of  variables $\mathtt{a}$ and $\mathtt{b}$ after the loop in Fig.~\ref{exprogram} terminates. Since we are in the top-level, we  simplify the final expression for $\mathtt{a''}$ and $\mathtt{b''}$ using interval arithmetics by replacing the counter variables with their corresponding bounds.
 Let us first express $\mathtt{b''}$ in terms of input variables.   
 \[
\begin{array}{lll}
  \mathtt{b''=}&\mathtt{b' - \frac{1}{2}* k_2 (-2*a' + k_2 + 2*k_1 - 1)} & ~ \\
  ~=&\mathtt{b' - \frac{1}{2}*a (-a + 2*k_1 - 1)}&~ (\mathtt{a'=a,a''=a'-k_2,a'' \le 0, k_2 \leq a' \rightarrow k_2=a}) \\

    ~=&\mathtt{b' + \frac{1}{2}*a (a - 2*k_1 + 1)}&~  \\
    ~=&\mathtt{b' + \frac{1}{2}*a (a - 2*[0,1] + 1)}&~  (\mathtt{k_1 \in [0,a''+1], a''=a-k_2,k_2=a  \rightarrow k_1 \in [0,1]}) \\
        ~=&\mathtt{b' + \frac{1}{2}*a ([a -1,a+1])}&~   \\
      ~=&\mathtt{[0,b] +  [\frac{1}{2}*a(a -1),\frac{1}{2}*a(a+1)]}&~  (\mathtt{b'=b-k_3, k_3 \in [0,b]\rightarrow b' \in [0,b]}) \\
            ~=&\mathtt{ [\frac{1}{2}*a(a -1),b+\frac{1}{2}*a(a+1)]}&~   \\

\end{array}
 \]
 Thus, we have
   $\mathtt{\frac{1}{2}*a (a - 1) \leq b'' \leq b + \frac{1}{2}*a (a + 1)}$ when the program terminates. Similarly, we obtain  $ \mathtt{a'' =0}$ as we have $\mathtt{a'=a,k_2=a,}$ $\mathtt{a''=a'-k_2.}$ 
   %

%
%
%
%
%
%
%
%

 % end input ./workout_ex.tex
 
%
% start input ./conclusion.tex
\section{Discussion and future work} \label{sec:conclusion}

We presented an approach to  solving numerical linear loops with linear or polynomial assignments, in which a central concept is that of a path program, to transform 
(i) recurrences with multiple recursive cases  into recurrences with only single recursive case, and  
(ii)  recurrences with multiple arguments into a set of recurrences for functions having only a single argument. 
Our approach, besides dealing with these, ensures that all loops in the resulting program are directly recursive 
(no mutual recursion) and
recurrences are expressed in terms of an induction variable $k$ (the path counter) allowing the recurrences to be solved and focussing the analysis problem on  finding bounds for $k$.
We suggested that it can be bounded from above by  the value of a ranking function of the corresponding loop. 
These transformations help to overcome typical limitations of CASs, since recurrences for  multi-case, multi-argument 
functions are usually not syntactically supported by these tools. But the success of our method depends on external tools such as ranking function synthesisers and CASs. 
As another limitation, our approach takes a multi-path loop and turns it into nested single-path loops. 
The result of such a transformation using regular expressions is linear in the number of paths,
since our approach produces one path predicate for each distinct subexpression
of the regular path expression \cite{big-small-step-vpt2020-short}, and the number of distinct subexpressions
in the transformed expressions is linear in the number of paths.
However,  transforming a given program into a multi-path loop expression of
form $(e_1 + \ldots + e_n)^*$ could cause an exponential blow-up in size; for example, the number of paths through a 
loop consisting of a sequence of conditionals is exponential in the number of conditionals.
This may prove to be a bottleneck in real applications. Finally, the transformations are limited to linear clauses 
since they rely on specific properties of regular expressions, but techniques for linearisation of clauses can sometimes 
be applied.

The approach seems promising though much work remains, both theoretical and experimental. 
Our immediate task is experimentation with a first implementation of the transformations, and integration
into an existing resource analysis tool \cite{ciaopp-sas03-journal-scp-short}.
Embedding this approach inside a recurrence solver such as Mathematica \cite{mathematica} could also
provide several advantages:  additional simplification and pre-solving, early detection of patterns, extrapolation and their propagation, and access to an efficient implementation. 

With regard to the removal of multi-path loops through transformations based on a regular path expression,
we have made some preliminary investigations on choosing good regular expression transformations.
There are several  (if not an infinite number) of valid equivalent transformations that induce different path programs.
The choice of such expression can affect precision of the analysis (though not its soundness). 
The role of lexicographical ranking functions for multi-path loops seems to be a promising aspect to
investigate.  For instance, for a loop whose regular expression is $(a+b)^*$, representing a loop with two paths $a$ and $b$,
suppose there is a lexicographical ranking function $\langle r_a,r_b \rangle$, where $r_a$ and $r_b$ are the ranking
functions for paths $a$ and $b$ respectively.  That is, the sequence of pairs $\langle r_a(x_1),r_b(x_1) \rangle,
\langle r_a(x_2),r_b(x_2) \rangle, \ldots$ is lexicographically ordered, where $x_1, x_2,\ldots$ is the sequence of
values of loop variables in successive loop iterations.
In this case, the replacement expression $b^* (ab^*)^*$ seems better than the equivalent $a^* (ba^*)^*$.
We might also compute a refined expression (possibly not  equivalent to the original)  that produces sound results
(preserving the feasible paths) but is more
precise than the original. 
The latter has been studied in \cite{Cyphertpopl19}, and we believe that it  is closely related to control-flow refinement 
for resource and termination analysis \cite{DomenechGG19} or program specialisation, as performed in polyvariant analyzers~\cite{mcctr-fixpt,spec-jlp}.

With regard to elimination of multi-argument functions in recurrences, replacing them by single-argument functions
of a loop counter, the key issue to investigate is how to improve the bounds on the counter's value.
We are using a ranking function, but this could be combined with other techniques, such as deriving invariants from 
the whole program, to
improve the bounds.  Interval arithmetic, which we used here, in general also loses precision, 
but could be combined with other
analyses.  The form of the chosen regular path expression can also play a role in the analysis of the bounds.  
For example, the path expression for a loop contains an $\epsilon$ expression (an empty path) as the ``base case".  
Thus the encoding loses the connection between the loop and its original base case, which is moved to another path.
Other regular expressions for loops could address this issue.

Extending our approach to extract and solve recurrence inequations as in \cite{farzanfmcad15} is another avenue for future work.  We are also interested in extending the current work to handle non-linear recursions either natively (possibly transforming context free grammars) or through iterative linearisation of the source clauses as in \cite{Kaflehcvs16}.

%
%
%
%
 % end input ./conclusion.tex
 
\textbf{Acknowledgements.} We thank the anonymous reviewers for their constructive comments and for bringing related work to our attention. 
Research partially funded by MICINN PID2019-108528RB-C21
\emph{ProCode} project,
the Madrid P2018/TCS-4339 \emph{BLOQUES-CM} program and the Tezos
foundation.

\bibliographystyle{eptcs}

\end{document}